\documentstyle[12pt]{article}
\begin{document}
\baselineskip=12pt

\title{Particle Dark Matter}

\author{David Spergel\\
Department of Astrophysical Sciences, Princeton University\\
 \& Department of Astronomy, University of Maryland, College Park}
\maketitle

\section{Introduction:
Three Arguments for\newline 
Non-baryonic Dark Matter}

Several lines of evidence suggest that some of the dark matter may
be non-baryonic: the non-detection of various plausible baryonic
candidates for dark matter inferred, e.g., from galaxy rotation curves
and from cluster of galaxy velocity dispersions, 
the need for non-baryonic dark matter for
theoretical models of galaxy formation, 
 and the large discrepancy between
dynamical measurements implying $\Omega_0 > 0.2$ and
the baryon abundance inferred from big bang nucleosynthesis,
$\Omega_b h^2 = 0.015$.  There are a number of well-motivated
dark matter candidates: massive neutrinos, supersymmetric
dark matter and ``invisible'' axions.  Many of these dark matter
candidates are potentially detectable by the current generation
of dark matter experiments.

\section{The Case For Non-Baryonic Matter}

While there is a consensus in the astronomical community
that most of the mass of our Galaxy and of most galaxies
is in the form of some non-luminous matter \cite{Trimble87},
there is only speculation
about its nature.

In his lecture, Charles Alcock (see the contribution by C. Alcock to
these proceedings) presents a report
of recent progress in efforts to detect baryonic
dark matter.  Here, I will focus on non-baryonic dark matter.

I will begin by presenting three arguments that suggest
that the dark matter is non-baryonic. None of these
arguments are definitive. John Bahcall has urged
the speakers to identify interesting problems
for graduate students. In addition to the
grand challenge of detecting the dark matter,
I believe that an easier problem is to
make some of the arguments for dark matter more compelling.

\subsection{We've looked for baryonic dark matter and failed}

Astronomers have already eliminated a number
of plausible candidates for the dark matter.
X-ray observations of galaxies imply that only a small fraction
of the mass of a typical galaxy is in the form
of hot gas \cite{Awaki94,Mulchaey93}. Even in rich clusters, hot gas makes up
less than 20\% of the total mass of the system \cite{Boute96}.
Neutral hydrogen gas is detectable through its 21 centimeter emission:
in most  galaxies, neutral gas comprises only 1\% of the mass
of the system \cite{Scodeggio93} and in only a handful of dwarf galaxies
does the neutral gas mass exceed the stellar mass.
 Even in these systems (e.g., DDO 240 \cite{Carrignan}),
neutral gas does not account for more than 20\% of the system mass.
Molecular gas is detectable through dipole emission of CO
and other non-homopolar molecules: in most galaxies, the
molecular gas mass appears to be less than the neutral
gas mass. Low luminosity (low mass) stars, M~dwarfs, have often been
proposed as a dark matter candidate but HST observations show that
faint red stars contribute less than 6\% of the unseen matter in the
galactic halo \cite{JBahcall94}.

If the dark matter is composed of baryons, then these
baryons must be clumped into dense bound objects to evade
detection.  Gerhard and Silk \cite{Gerhard95} have proposed that
the dark matter consists mostly of very dense
tiny clouds of molecular gas. Their model, while
provocative, is only marginally consistent with
current observational limits. A more widely
accepted proposal is that the dark matter consists
of very low mass stars, called brown dwarfs. These brown
dwarfs are not massive enough to burn hydrogen, so
that their only energy source is gravitational energy.

While these brown dwarfs are difficult to detect
through their own emission, they are
potentially detectable through the gravitational 
effects.  Paczynski \cite{Paczynski86} proposed gravitational lensing
searches for these objects. Several
groups have begun searching for these events
in an effort to probe the nature of the dark matter.

So far, MACHO searches are not finding as many events
as predicted by spherical halo models \cite{Alcock95}; however,
they can not yet rule out MACHOs as the dominant component
of the halo. The current experiment is limited
by both small number statistics and by
uncertainties in galactic parameters. Many important
galactic parameters such as the circular speed, disk scale length and
the local surface density are still quite uncertain.
Because of these uncertainties, the local
halo density is not certain to a factor of two.

It is particularly important to accurately determine
the local circular speed as our estimates of the local dark matter
density is very sensitive to its value:
$${\partial \log \rho_{\rm halo} \over \partial \log v_c} = 2
{v^2_{\rm tot}
\over v^2_{\rm tot} - v^2_{\rm disk}} \sim 4\ .$$
(Deriving this formula is a good exercise for a student
new to dynamics.  For an excellent introduction
to the subject, see Binney \& Tremaine \cite{Binney}).
Thus, a 10\% uncertainty in local circular speed translates
into a 40\% uncertainty in the local dark matter density.
Without more accurate determinations of $v_c$, it is
difficult to definitively argue that MACHOs can
not comprise much, if not all, of the mass of the dark halo.

There is also a need for better models  of the LMC and more accurate
measurements of its properties.  Some of the lensing
events reported by the MACHO and EROS collaborations
may be due to "self-lensing" by the LMC \cite{Sahu94}
rather than dark matter in the halo.

\subsection
{We can't seem to make large scale structure without WIMPs}

All of the most successful models for
forming large scale structure assume
that most of the universe is composed of cold
dark matter.

Models in which the primordial fluctuations are
adiabatic and the universe is comprised only of
baryons and photons are
ruled out by CBR observations. The predicted level
of fluctuations in these models exceed the observed level
by more than an order of magnitude.
Isocurvature models \cite{Peebles90} fare better; however,
these models also appear to
be in conflict with CBR observations \cite{Chiba94}.

The current 
``best fit'' models have $\Omega_0 \simeq 0.3$,
$H_0 \simeq 0.75$, $\Omega_b\simeq 0.03$ and either
a cosmological constant or space curvature
(see Steinhardt's talk in these proceedings for a review). These models
fit COBE observations; are
consistent with age and $H_0$ determinations; 
are consistent with LSS power spectrum, and
are consistent with most large scale velocity measurements.
While they are in conflict with the large velocities
detected by Lauer \& Postman \cite{Lauer}, these large
velocities are controversial \cite{Kirshner95}.
Numerical simulations suggest that these models also
agree with the properties of rich clusters \cite{Bahcall94}.

Despite the success of structure formation
models that assume non-baryonic dark matter,
no one has proven a ``no-go'' theorem that
rules out baryon-only models. It is an
interesting challenge to determine what
observations are needed to rule out
these models.

\subsection{Dynamical Mass is Much Larger than Big Bang
Nucleosynthesis Allows}

Measurements of the mass-to-light ratios in
clusters suggest that $\Omega_{\rm tot}$, the ratio
of the total density of the universe to
the critical density, exceeds 0.2 \cite{Bahcall95}. This determination
of $\Omega_{\rm tot}$ is consistent with measurements
based upon the large-scale
velocity fields and the dynamics of
the large-scale structure \cite{Strauss95}.
Values of $\Omega$ less than 0.2
are very difficult to reconcile with the
500 km/s random velocities seen in large
scale structure surveys and even harder to reconcile with
large-scale streaming motions.

The observed (presumed cosmological) abundances of deuterium, helium and 
lithium
are only consistent with standard big bang nucleosynthesis
if the baryon density is much less than $\Omega_{\rm tot}$.
The best fit value for
$\Omega_b h^2 \simeq 0.015$, which is nearly
an order of magnitude below the dynamical
values \cite{Walker91}.
For example, if $H_0 = 75$ km/s/Mpc, $\Omega_b =0.2$ implies
that $Y$, the Helium/Hydrogen abundance
ratio, is $0.262$ and $D/H$, the Deuterium/Hydrogen
abundance ratio, is $10^{-6}$ \cite{Walker91}
while if
$H_0 = 50$ km/s/Mpc, $\Omega_b =0.2 $ implies
$Y = 0.253$ and $D/H = 5 \times 10^{-6}$.
There are many extragalactic HII regions with $Y < 0.25$
and best estimates imply $Y \simeq 0.24$. These
observations appear to require either a significant
modification of our ideas about big bang
nucleosynthesis or the existence of copious
amounts of non-baryonic dark matter. (See, however,
Goldwirth \& Sasselov \cite{Goldwirth95} for a dissenting view).

All of the proposed 
modifications of BBN appear to violate known
observational constraints.
For example, Gnedin \& Ostriker \cite{Gnedin91} proposed
that an early gamma-ray background
photodissociated some
of the primordial Helium. This model predicts a spectral
distortion of $y > 7\times 10^{-5}$ and a fully
ionized universe. $y$ describes the deviation of the
observed spectrum from the thermal spectrum and
is a measure of the energy injection in the early universe.
COBE \cite{Mather92} found that the observed spectrum was consistent
(within the experimental errors) with a thermal
spectrum and constrained  $y < 2.5 \times 10^{-5}$.

Inhomogeneous nucleosynthesis models have been studied
extensively in the past few years. However, Thomas
et al. \cite{Thomas94} found that 
even models with large
inhomogeneities imply $Y > 0.25$ for $\Omega_b h^2 > 0.05$.
Thus, they are also not consistent with $\Omega_b = \Omega_{\rm tot} = 0.2$.

While the theory of big bang nucleosynthesis is well developed,
there is still uncertainty in converting the observed
line ratios to abundances.  Most of the abundances for
external systems assume a spherical clouds with
constant rates of ionization.  It would be interesting
to study a nearby system such as the Orion nebula
and estimate the error associated with this
approximation in the analysis.  Goldwirth \& Sasselov \cite{Goldwirth95}
have made an important first step in studying the sensitivity
of these element abundances to model uncertainties.
There is a need for more work.

While none of these three arguments is incontrovertible,
they all do suggest that most 
of the universe is in non-baryonic matter.
The rest of this paper will review the most popular
proposed candidates for non-baryonic dark matter
and consider various schemes for detecting its presence.

\section{Neutrinos as Dark Matter}

In the standard big bang model, copious numbers
of neutrinos were produced in the early universe.
The universe today is thought to be filled
with 1.7 K thermal neutrino radiation, the neutrino
complement to the thermal radiation background.
If these neutrinos are massive, then they
can make a significant contribution
to the total energy density of the universe:
\begin{equation}
\Omega_\nu h^2 \simeq \left({m_\nu \over 100 {\rm eV}}\right)
\end{equation}

Recent results from solar neutrino experiments have revived
interest in neutrinos as dark matter candidates. As John
Bahcall has described in his talk (see these proceedings), recent experiments 
appear
to be consistent with the MSW solution to the solar neutrino deficit.
The MSW solution implies that the difference in
mass squared between the electron neutrino and another
neutrino family is of order $10^{-5}$ eV$^2$. While
this mass difference is much smaller than the mass
needed for neutrinos to be the dark matter, it does
suggest that neutrinos are massive. It is thus certainly
possible that the MSW effect is due to oscillations
between electron and mu neutrinos and
that the tau neutrino is much more massive and
comprises much of the dark matter.

There are several astronomical problems
for neutrino dark matter models.
Because cosmic background
neutrinos have a Fermi-Dirac distribution, they have a maximum
phase-space density, which implies a maximum
space density \cite{Tremaine}.  Dwarf irregular galaxies \cite{Carrignan} have
very high dark matter densities and dwarf spheroidals
 \cite{Gerhard92} have even higher dark matter densities:
neutrinos can not be the dark matter in these systems.
So, if neutrinos are the dark matter in our Galaxy, then
there is a need for a second type of dark matter for low
mass galaxies
 \cite{Gerhard92}. Neutrino plus baryon models have a difficult time
forming galaxies early enough and these models predict
galaxy clustering properties significantly different
from those observed in our universe.

There are, however, several modified neutrino
models that appear more attractive.
Cosmological models in which cosmic string
seed fluctuations in the hot dark matter
have several promising features for structure
formation \cite{Albrecht92}.
Mixed dark matter models in which neutrinos
comprise 20\% of the dark matter and the
rest of the dark matter is comprised of cold
dark matter also appear to be consistent
with a number of observations of large scale
structure \cite{Primack95}.

\subsection{Detecting Massive Neutrinos}

While it is very difficult to detect
the cosmic background of neutrinos directly, there are
several experimental approaches that
might be able to measure the mass of the neutrino.
As I noted earlier, the detection of a stable several eV
neutrino would imply that neutrinos comprise
a significant fraction of the mass of the universe.

The classical approach to measuring neutrino
mass are measurements of the
$\beta$ decay endpoint. Current limits
from these experiments imply that the electron
neutrino is not the predominant component of the dark matter; however, these
experiments cannot place astrophysically
interesting constraints on the mass of the mu or tau neutrino.

If the neutrino is a Majorana particle, then it might
be indirectly detected through the detection of
a neutrinoless double beta decay. Deep underground
experiments looking for rare decays have
placed very interesting limits \cite{Baylsh94} on the electron neutrino
mass:
$m_{\nu_e} < 0.68$ eV. 
This is a limit on massive neutrinos
{\it if} the most massive eigenstate contains
a significant fraction of the electron flavor eigenstate
and does not apply to all neutro models.

Neutrino oscillation experiments are sensitive
to mass differences, usually
$\Delta m^2 = m_{\nu_\mu}^2 - m_{\nu_e}^2$ and sometime
$m_{\nu_\tau}^2 - m_{\nu_e}^2$. Recent results from
the Los Alamos experiment
 \cite{Athanassopoulos95}, which suggest a detection
of neutrino oscillations, are controversial \cite{Hill95}. 

There is a possibility of an astronomical detection
of neutrino mass using neutrinos from a supernova explosion. 
If the neutrinos are massive, then more-energetic neutrinos
arrive earlier than less-energetic neutrinos. Thus,
neutrino detectors would first see higher energy events
and then see less energetic events. This effect was
not observed in SN 1987A, which suggests
that $m_{\nu_e} < 15$ eV \cite{Bahcall}. Observations of a galactic
supernova by Sudbury detector, which is sensitive 
to $\nu_\mu, \nu_\tau$ could place interesting limits
on their masses and possibly rule out
neutrinos as cosmologically interesting.

\section{WIMPs}

There is broad class of particle physics candidates
for the dark matter that are referred to
as Weakly Interacting Massive Particles or WIMPs.
This class includes several proposed particles
(massive Dirac neutrinos, cosmions, SUSY relics)
that have masses of order a few GeV to 
a few hundred GeV and interact
through the exchange of W's, Z's, higgs bosons
and other intermediaries. In this
talk, I will give a brief introduction to
WIMPs. I refer interested readers to  recent, 
more detailed reviews \cite{Smith90,Primack88,Jungman95}.

The early universe is a wonderful particle accelerator.
WIMPs could be produced through reactions such as
$e^+\ e^- \to X \bar X$, where $X$ denotes
the WIMP particle. WIMPs, of course, can be
annihilated through the backreaction,
$X \bar X \to e^+\ e^-$.
As long
as $T > m_X$, the WIMP number density would
be comparable to the number density of
electrons, positrons, and photons. However,
once the temperature drops below $m_X$, the WIMP
abundance begins to drop. It will fall until
the WIMP number density is so low that
the WIMP mean free time for annihilation exceeds
the age of the universe. This ``freeze-out'' occurs
at a density determined by the WIMP annihilation cross-section
and implies that
$$\Omega_{x} h^2 \simeq \left({\sigma_{\rm ann}\over 10^{-37} cm^2}
\right)^{-1} $$

The first proposed WIMP candidates were heavy fourth
generation neutrinos \cite{Hut,Lee}. 
If the neutrino mass was of order 2 GeV, then
its relic abundance would be sufficient for $\Omega_{\nu} =1$.
Experimental dark matter searches \cite{Ahlen87} ruled
out these particles as dark matter candidates.

Supersymmetry is an elegant extension
of the standard model of particle physics.
It is the only so-far ``unused'' symmetry of the Poincare group
and has the virtue of protecting the weak
scale against radiative corrections from GUT and Planck scale.
Local supersymmetry appears to be an attractive route
towards unifying all four forces and is a basic ingredient
in superstring theory. Supersymmetry transforms
bosons into fermions (and vice-versa). As supersymmetry
has a new symmetry, R parity, it can imply the
existence of a new stable particle.
In much of the parameter space of the minimal
supersymmetric model, this new stable particle
(which we will refer to as the``neutralino'') has
predicted properties such that it would comprise
much of the density of the universe \cite{Ellis84}.

\subsection{Searching for WIMPs}

While WIMPs interact weakly, they are potentially
detectable \cite{Goodman85,Wasserman86,Jungman95}.
The flux of WIMPs through an
experiment is quite large:
10$^6$ (m/GeV)$^{-1}$ cm$^{-2}$ s$^{-1}$. The
difficulty lies in detecting the rare WIMP
interactions with ordinary matter.

The challenge for dark matter experimenters
is to design an experiment that is simultaneously
sensitive to few keV energy depositions and has
a large mass (many kilograms) of detector material.
The experiment must also have 
superb background rejection as the expected
event rate, less than an event/kilogram/day, is
far below most backgrounds. There are two 
potentially experimental signatures that can
aid in the WIMP search: a roughly 10\% annual
modulation of the event rate due to the Earth's
motion around the Sun \cite{Drukier86}
and a large ($\sim 50\%$) asymmetry in
the direction of the WIMP flux due to the
Sun's motion through the galactic halo \cite{Spergel88}.

The first generation of WIMP experiments were
rare-event experiments that were adapted to
search for dark matter. The first set of experiments
were  ultra-low background
germanium semiconductor experiments \cite{Ahlen87,Caldwell88,Reusser91}
that
were developed as 
 double beta-decay experiments and modified into
dark matter detectors. In these experiments,
a recoiling Ge nucleus produces $e^-$-hole pairs
that are detectable down to recoil energies $\sim 5$ keV.
These experiments have been limited by microphonics,
electronic noise, and by cosmogonic radioactivity.

We are now entering the era of second generation 
experiments that have been designed primarily as
dark matter detectors.   In this section, I will
highlight several of the promising experimental
technologies.

The Heidelberg-Moscow
germanium experiment is
a modification
of the early germanium experiments. It consists of
6 kilogram of purified $^{76}$Ge detector in Gran Sasso Tunnel.  
Since it does not contain $^{68}$Ge, it has a reduced cosmogonic
background. In this experiment,
electronics and microphonics are the dominant background.
This experiment places the best current limits
on the halo density of WIMPs more massive than 50 GeV \cite{Beck94}.

Rather than detecting the electron-hole pairs produced
by recoiling nuclei, the Stanford silicon experiment \cite{Young90}
detected the ballistic phonons produced by recoiling
silicon nuclei. This experiment has been
calibrated by neutron bombardment. The Munich
group is developing a silicon detector that
will detect the ballistic photons with an SIS junction
 \cite{Peterreins91}.

At Berkeley, the CfPA group is developing a
detector that is sensitive to 
both phonon and electron-hole pairs. This 
dual detection allows much better background rejection as
electrons excited by radioactive decays have a different
photon and electron-hole pair signature than nuclear recoils.
Neutron bombardment experiments suggest that this dual
detection technique can reject
$\sim 99\%$ of radioactive background \cite{Shutt92}.
A more massive experiment that utilizes this
technique has the potential to probe into
interesting region of parameter space in supersymmetric
theories.

Several groups are developing scintillators that
are potential WIMP detectors. There
are several scintillator experiments currently
under development:
a 36.5 kg NaI experiment in Osaka that has 
begun to place interesting limits on
heavy neutrinos \cite{Fushimi93,Ejiri93};
the Rome/Beijing/Saclay experiment \cite{Bottino92}, a smaller detector,
with sensitivities similar to the Osaka experiment;
and a Munich sapphire scintillator experiment
that is designed to be sensitive to
low mass ($m < 10$ GeV) WIMPs.
This technology has several advantages
over the germanium and silicon semiconductors;
the material is 
sensitive to spin-dependent coupling (although,
this is now thought to be less important
for supersymmetric dark matter detection \cite{Jungman95})
and it is relatively easy to build very large mass
detectors.
The challenge for these experiments is 
to improve their background rejection. Spooner
\& Smith \cite{Spooner93} suggest that it might be possible
to have 
some rejection of radioactive $\gamma$'s in these
NaI scintillators 
through measurements of UV and VIS signatures of recoils
 \cite{Spooner93}.
\subsubsection{Gas Detectors}

Time-Projection Chamber (TPC)  detectors have been used extensively in particle
physics experiments. While a gas detector with
sufficient mass to be sensitive to neutralinos
would have an enormous volume, this technology
does offer the possibility of detecting the
direction of WIMP recoil. Due to the Earth's
motion around the Sun, the WIMP recoil events are
expected to be highly asymmetric \cite{Spergel88}.
Buckland et al. \cite{Buckland94} report their development
of a 50 g H prototype detector. This
detector, developed at UCSD, has been tested with
neutron source and is potentially scalable
to larger masses.

\subsubsection{Superconducting Grains}

Superconducting grains have an illustrious history
in dark matter detection.  Drukier \& Stodolsky \cite{Drukier84}
proposed superconducting grains for neutrino detectors and
this work led Goodman \& Witten \cite{Goodman85} and Wasserman 
\cite{Wasserman86}
to propose the development of WIMP detectors.

A superconducting grain detector would consist of
numerous micron size superconducting grains in a meta-stable
state.When one of these grains is heated by WIMP recoil, it would
undergo a phase transition to the normal state. The resultant change
in $B$ field would be detected by a SQUID. Most background
events, due to radioactivity, would flip multiple grains in the 
detector. Since the events can also be localized in the detector,
this can further enhance background rejection as
background events should occur primarily near the outside
of the detector.
The challenge for  superconducting grain detector development is the 
production of large number of high quality grains.
Recently, the Bern group \cite{Abplanalp94} has been able to report significant
progress in this direction: they
been able to build a superconducting grain detector
with several different types of grains (Sn, Al and Zn grains),
which they have calibrated with a neutron source.

\subsubsection{``Old'' Mica}

WIMP detection requires exposure
times of $\sim 100$ kg-years. A novel approach
is to replace the 
100 kg detector with small amounts of material that
has been exposed for nearly a billion years.
Snowden-Ifft and collaborators \cite{Snowden95} have looked
for tracks produced by WIMP scatters off of
heavy nuclei (such as cadmium) in ancient Mica.
They identify these tracks by etching the Mica
and have calibrated their experiment by
bombarding the Mica with a neutron source.

\subsubsection{Atomic Detectors}

Recently, Glenn Starkman and I proposed
searching for inelastic collisions of SUSY relics
with atoms \cite{Starkman95}. 
The cross-sections for these
interactions are largest for $\delta E \sim 1$ eV. 
While the cross-section for atomic interactions
are smaller than nuclear interactions, there
is a wider range of material that could
be used for detecting these atomic interactions.
There are not yet any experimental schemes
proposed to look for WIMP-atom scatterings.  This
proposal requires more experimental and theoretical study.

\subsection{Indirect WIMP detection}

The Sun can potentially serve as an enormous WIMP
detector. WIMPs streaming through the galactic
halo would be gravitationally focused into the
Sun, where they would be captured through
collisions with atoms in the Sun's center \cite{Press85}.
Neutralinos are their
own anti-particles; thus, the neutralinos
in the Sun would annihilate each other. 
When neutralinos annihilate, they will produce
high energy neutrinos that are potentially
detectable in terrestrial experiments \cite{Silk84}.
These few GeV neutrinos are
much more energetic than the MeV solar
neutrinos produced through solar nucleosynthesis.
There is also the possibility of detecting
WIMPs in the halo through their annihilation
into protons and anti-protons, into electrons
and positrons and into $\gamma$'s.  The predicted rates for
these processes are unfortunately rather low \cite{Diehl95}.

There have been several experiments that
have looked for WIMP annihilations in the Sun.
Currently, there are limits from the 
Kamionkande, Frejus, and MACRO experiments.
In the coming years, we can look forward to
more sensitive searches by the 
DUMAND, AMANDA and NESTOR experiments.
While these searches are worthwhile,
Kamionkowski et al. \cite{Kamionkowski95} have argued that direct
experimental searches may be a more effective
technique than searches for neutrinos from
annihilations of SUSY relics in the Sun. However, for
the rarer models with predominantly spin interactions,
the converse is most likely true
They conclude that for most of parameter space,
1 kg of direct detector is equivalent to 10$^5$ -10$^7$ m$^2$ of indirect
detector.

\subsection{What is to be Done?}

Besides the challenge of helping to make any of the promising
experiments discussed above work, there are a number
of interesting open problems in the WIMP detection field for
both theorists and experimentalists.
Advances at LEP and at the Tevatron continue to place new
limits on the properties of SUSY particles and may
provide hints of their existence.  We need an  on-going reassessment
of the viability of different experiment approaches
(see e.g., \cite{Kamionkowski95}).  There is still
much work to be done on the interactions of neutralinos
with ordinary matter (see e.g., \cite{Starkman95}).  In particular,
it would be useful to consider the excitation of atomic levels
through WIMP-nuclei collisions.

Advances in technology may enable new kinds of WIMP detectors.
It would be very exciting to be able to build a detectors
composed large numbers ($\sim 10^{31}$) spin aligned nuclei.
As this detector would have directional sensitivity, it would be
sensitive to the large angular asymmetry in the WIMP flux \cite{Spergel88}
The development of new purification techniques in the semi-conductor
industry may help facilitate the construction of ultra-low background
Silicon and Germanium detectors.  It would be very exciting
if an experiment such as DUMAND or AMANDA with their
large detection volumes could be redesigned so that
it was sensitive to SUSY relics scattering events.  Because
of their large active volumes, even lower event rate
processes such as inelastic scattering are of potential
interest for these experiments.
Close collaborations between experimentalists, theorists and
technologists are need to advance the search for SUSY relics.

\section{Axions}

Axions are another well-motivated dark matter candidate.
While axions are much lighter than the SUSY relics discussed
in the previous section and are produced by a very different
mechanism, they are indistinguishable to theoretical
cosmologists studying galaxy formation and the origin
of large scale structure. Both axions and SUSY relics
behave as cold dark matter (CDM) and cluster effectively
to form galaxies and large-scale structure.  (See
Steinhardt's and Ostriker's articles on structure formation).

Axions were proposed to explain the lack of CP
violation in the strong interaction \cite{Weinberg,Wilczek}.
They are associated with a new U(1) symmetry:
the Peccei-Quinn symmetry \cite{PQ}. 
As originally proposed, axions interacted strongly
with matter. When experimental searches failed to
detect axions, new models were proposed that
evaded experimental limits and had the interesting
consequence of predicting a potential dark matter
candidate \cite{Kim,SVZ,Zhitnitsky,DFS}.

In the early universe, axions can be produced
through two very distinct mechanisms. At the
QCD phase transition, the transition at which free quarks
where bound into hadrons, a bose condensate of
axions form and these very cold particles
would naturally behave as cold dark matter.
Axions can also be produced through the decay of strings
formed at the Peccei-Quinn phase transition \cite{Davis86,Davis89}.
Unless inflation occurs after the P-Q phase transition,
string emission is 
thought the dominant mechanism for axion production.
While Sikivie and collaborators \cite{Hagmann91}
has argued that Davis and Shellard
overestimated string axion production, recent analysis \cite{Battye95}
confirm that strings are likely to be the dominant
source of axions.  Axionic
strings will not produce an interesting level
of density fluctuations as their predicted
mass per unit length is far too small to be
cosmologically interesting.

The properties of the axion are basically set
by its mass, $m_a$, which is inversely proportional
to the scale of Peccei-Quinn symmetry breaking, $f_a$.
The smaller the axion mass, the more weakly the axion is coupled
to protons and electrons.  Raffelt \cite{Raffelt95} reviews the
astrophysical arguments that imply $m_a < 10^{-2}$ eV. If
the axion had a larger mass, then it
would have had observable effects on stellar evolution
and on the dynamics of SN 1987A.
If we require that the energy density in axions
not ``overclose'' the universe, then
$\Omega_a h^2 < 1$ implies that $m_a > 1 \mu$eV.
If strings play an important role in axion production,
then the cosmological limit lies closer to $m_a > 1 meV$
and there is only a narrow window for the axion model \cite{Raffelt95}.

Axions are potentially detectable through their
weak coupling to electromagnetism \cite{Sikivie83}.
In the presence of a strong magnetic field,
the axionic dark matter could
resonantly decay into two photons.
The first generation of detectors consisted
of experiments in Florida  and at BNL 
that looked for this decay in a tunable resonant
cavity. Since the Peccei-Quinn scale is not well determined,
these experiments have to scan a wide range of frequencies
in their search for the axion. These experiments
were an important first step towards probing an interesting
region of parameter space.

In the past few years, the search for axions has been
revived by two new experimental efforts. Karl
von Bibber \cite{vonBibber94} and his group at LLNL have built a
cryogenically cooled cavity; this
detector should be able to reach into cosmologically interesting 
region of parameter space.In Kyoto, Matsuki \cite{Matsuki94} and
his group plan to use an atomic beam of 
Rydberg atoms as an axion detector. This detector
would detect an axion in the galactic halo through
its excitation of a Rydberg atom in the n-th energy
state to the n+1 energy state. The Kyoto collaboration
also promises to probe the cosmologically interesting
region of parameter space.

\section{Conclusions}

While there is no conclusive evidence for non-baryonic dark matter, 
there are strong hints that it may comprise most of the mass
of the universe.
There are several well motivated particle physics
candidates for non-baryonic dark matter.
Most excitingly, these
candidates are potentially detectable in experiments
currently under development.

\section*{Acknowledgments}
I would like to thank John Bahcall, Marc Kamionkowski. 
Chris Kolda, \& Bill Press for comments on an earlier version.  I would
also like to thank Bernard Sadoulet and Karl von Bibber
for loaning me slides and updating me on recent experimental
progress.

\end{document}